\documentclass[12pt]{iopart}

\usepackage{iopams}
\usepackage{amsthm}
\usepackage{graphicx}

\usepackage{bm}
\usepackage{hyperref}
\usepackage{graphics}
\usepackage{slashed}
\usepackage{amssymb}
\usepackage{url}
\usepackage[usenames,dvipsnames,svgnames,table]{xcolor}
\usepackage[T1]{fontenc}

\makeatletter
\newcommand{\tpitchfork}{%
  \vbox{
    \baselineskip\z@skip
    \lineskip-.52ex
    \lineskiplimit\maxdimen
    \m@th
    \ialign{##\crcr\hidewidth\smash{$-$}\hidewidth\crcr$\pitchfork$\crcr}
  }%
}
\makeatother

\usepackage{color}

\begin{document}

\title{Mimicking a rotating black hole with nonlinear electrodynamics}
\author{\'Erico Goulart${}^{1}$ and Eduardo Bittencourt${}^{2}$\footnote{Corresponding author}}
\address{${}^{1}$Federal University of S\~ao Jo\~ao d'el-Rei, C.A.P. Rod.: MG 443, KM 7, CEP-36420-000, Ouro Branco, MG, Brazil}
\address{${}^{2}$Federal University of Itajub\'a, Itajub\'a, Minas Gerais 37500-903, Brazil}
\ead{egoulart@ufsj.edu.br, bittencourt@unifei.edu.br}

\begin{abstract}
We exhibit the first analogue model of a rotating black hole constructed in the framework of nonlinear electrodynamics. The background electromagnetic field is assumed to be algebraically special and adapted to a geodesic shear-free congruence of null rays in Minkowski spacetime, the Kerr congruence. The corresponding optical metric has a Kerr-Schild form and, it is shown to be characterized by three parameters, thus predicting the existence of an ergosurface, a horizon, and a slice identical to one also present in the Kerr metric.
\end{abstract}

\vspace{2pc}

%
%
%

\section{\label{sec:level1} Introduction}

The Kerr spacetime, an exact solution to Einstein's field equations, describes spacetime around a rotating and axially symmetric massive object. Rotation is a common characteristic of astronomical objects, from stars and black holes to entire galaxies. Hence, since its discovery by Roy Kerr in 1963 \cite{Kerr_1963,kerr1965gravitational}, this solution has played a fundamental role in understanding gravitational phenomena in high-energy and intense gravity regimes and leads to the correct limits in the weak field approximation. However, the mathematical complexity of the solution makes exploring its physical implications challenging. In this context, analogue models of gravity offer a promising approach \cite{barcelo2005}.

In the last years, some analogue models of rotating black holes have been investigated (see reference \cite{novello2002artificial} for a review). Important examples include the geometry felt by phonons entrained in a nonrelativistic rotating fluid vortex \cite{visser2005vortex}, the relativistic condensate approach exposed in reference \cite{giacomelli2017rotating} and the model of a planar soliton moving in a superfluid $^{3}$He \cite{jacobson1998event}. More recently, in an effort to invariantly characterize the curvature structure of a class of analogue spacetimes, the Petrov types of a variety of geometries have been investigated in detail \cite{baak2023petrov}. In particular, the authors re-obtain the result that the equatorial constant-time slice of the Kerr geometry (under very specific engineered conditions), can be mimicked by a compressible vortex flow. From a broad perspective, the models provide qualitative discussions concerning effective horizons, ergoregions, superradiance, and the dragging effect. However, the incorporation of rotation into analogue spacetime descriptions still remains difficult and the study of more sophisticated models is very welcome.

In this article, we construct an analogue model of gravity based on nonlinear electrodynamics \cite{Novello2000,DELORENCI2000134,DUNNE_2005,breton2007nonlinear,Sorokin_2022,alam2022review}, mimicking the main features of the Kerr solution \cite{Chandra1983,Stephani_2003,visser2008kerr}. Specifically, our model is grounded on an algebraically special (null) electromagnetic bivector in Minkowski spacetime which is adapted to a geodesic shear-free congruence of null rays, the Kerr congruence. From the very beginning, this choice automatically captures the essential geometric structure of Kerr spacetime, which manifests itself in the corresponding optical metric. The latter is characterized by three parameters, one of them being related to the nonlinearity of the theory and the others related to the \textit{mass} and \textit{rotation} of the solution. Furthermore, the effective light cones are shown to be controlled by a class of scalar functions directly related to the intensity of the underlying background field. Similar functions already appear in the seminal work of Teukolsky on the electromagnetic perturbations around a rotating black hole \cite{teukolsky1973perturbations,Chandra1983} and have been generalized recently by \cite{pelykh2018class,pelykh2018null}.

Notably, our model exhibits an ergosurface, a horizon, a three-dimensional slice identical to the Kerr geometry, and an interesting singularity structure, a result unprecedented in analogue models made with nonlinear electrodynamics. It is our hope that the model enriches the catalog of known analogue models and provides an accessible platform for exploring extreme gravitational phenomena in a laboratory framework, such as energy extraction from rotating holes and orbital precession of particles.

The paper is summarized as follows: in section 2 we construct the optical metric for null electromagnetic solutions of nonlinear electrodynamics, emphasizing their Kerr-Schild type. In section 3 we present Mawxell's equations in the Newman-Penrose formalism. In section 4, we use the Kerr congruence to exhibit a corresponding null solution in the Kerr metric and, in section 5, we transpose it to the Minkowski spacetime also as a null solution. Finally, in section 6, we discuss the main features of the optical metric obtained from this solution, interpreting it as an analogue rotating black hole. Throughout the text, we set the speed of light to the unit.

\section{Optical metrics}

To begin with, we let $(\mathcal{M},\mathring{g}_{ab})$ denote a four-dimensional Minkowski spacetime with metric signature convention $(+,-,-,-)$ and write 
\begin{equation}\label{invs}
\psi=\frac{1}{2}F_{ab}F^{ab},\quad\quad\quad\phi=\frac{1}{2}\star F_{ab}F^{ab},
\end{equation}
for the two independent invariants of the electromagnetic field $F^{ab}(x)$, where the Hodge dual bivector reads as
\begin{equation}\label{dual}
\star F^{ab}=\frac{1}{2}\varepsilon^{ab}_{\phantom a\phantom a rs}F^{rs},
\end{equation}
with the Levi-Civita tensor given by $\varepsilon_{abcd}=\sqrt{-\mathring{g}}\ [abcd]$, $\mathring{g}$ being the determinant of the metric in any coordinate system and the totally anti-symmetric permutation symbol defined such that $[0123]\equiv +1$. 

For the sake of simplicity, we shall consider a gauge invariant nonlinear theory of electrodynamics in a source-free region of $(\mathcal{M},\mathring{g}_{ab})$ provided by the action\footnote{The inclusion of the invariant $\phi$ in the Lagrangian does not affect our main conclusions.}
\begin{equation}\label{action}
S=\int \mathcal{L}(\psi)\sqrt{-\mathring{g}}\ d^{4}x,
\end{equation}
where the Lagrangian density $\mathcal{L}(\psi)$ is an arbitrary function of the invariant $\psi$. Variation with respect to the four-potential $A_{a}$ yields the quasi-linear system of first-order partial differential equations
\begin{equation}\label{nled}
\left(\mathcal{L}_{\psi}F^{ab}\right)_{;b}=0,\quad\quad\quad \left(\star F^{ab}\right)_{;b}=0,
\end{equation}
with $\mathcal{L}_{\psi}\equiv \partial\mathcal{L}/\partial\psi$ for conciseness and $;$ standing for covariant derivative with respect to the background metric $\mathring{g}_{ab}$. A necessary (but not sufficient) condition for this theory to admit a well-posed Cauchy problem is that $\mathcal{L}_{\psi}\neq 0$, which we shall assume henceforth (see \cite{GoulartdeOliveiraCosta:2009pr,Abalos:2015gha,Goulart:2021uzr} for details).

It is well known that the \textit{optical metrics} governing the behavior of high frequency and low amplitude perturbations of the electromagnetic field are given by \cite{Goulart_2024}
\begin{equation}
\tilde{g}_{ab}=\mathring{g}_{ab}-\frac{\chi}{1-\chi\psi} F_{a c}F^{c}{}_{b},\quad\quad \tilde{g}^{ab}=\mathring{g}^{ab}+\chi F^{a}_{\phantom a c}F^{cb},
\end{equation}
where the quantity $\chi\equiv-2\mathcal{L}_{\psi\psi}/\mathcal{L}_{\psi}$ controls the magnitude of the nonlinearity and is assumed positive for the sake of causality \cite{Goulart_2024}, and the bivector $F^{ab}$ describes the specific structure of the background field. It turns out that if the latter is null, i.e. if both invariants vanish simultaneously, there exists a real null vector $n^{a}$ such that
\begin{equation}\label{PND}
F^{a}_{\phantom a b}n^{b}=0,\quad\quad\quad \star F^{a}_{\phantom a b}n^{b}=0.
\end{equation}
This vector is called a principal null direction of the electromagnetic field (PND) and the congruence spanned by it is denoted by $\Gamma(n)$. Interestingly, for a null background field adapted to $\Gamma(n)$, there follows
\begin{equation}\label{intensity}
F_{a c}F^{c}{}_{b}=E^{2}n_{a}n_{b},
\end{equation}
where the scalar $E(x)$ is the intensity of the field. Then, the optical metric and its inverse reduce to \cite{Goulart_2024}
\begin{equation}\label{effmets}
\tilde{g}_{ab}=\mathring{g}_{ab}- H_{{\rm opt}} n_{a}n_{b},\quad\quad\quad \tilde{g}^{ab}=\mathring{g}^{ab}+ H_{{\rm opt}} n^{a}n^{b},
\end{equation}
with $H_{{\rm opt}}\equiv \chi E^{2}$, for conciseness. Furthermore, since $\mathcal{L}_{\psi}$ is constant on top of the null field, the equations of motion (\ref{nled}) somehow linearize and the background is actually constrained to satisfy Maxwell equations in vacuum. 

In what follows, we shall evaluate the optical metrics engendered by null Maxwell fields adapted to the so-called Kerr congruence. Our main motivation is to see whether the corresponding optical metrics are able to mimic aspects of rotating black holes as described by general relativity. However, before we proceed, we review some aspects of Maxwell equations within the Newman-Penrose (NP) formalism in an
arbitrary spacetime.

\section{Maxwell equations in NP formalism}

In this section, we consider an arbitrary four-dimensional spacetime $(\mathcal{M},g_{ab})$ and form a tetrad of null vectors $z^{a}_{m}=\{l^{a},n^{a},m^{a},\bar{m}^{a}\}$, with the index $m=1,...,4$ labeling each leg of the tetrad. As usual, the tetrad basis will be normalized as follows:
\begin{equation}\label{norm}
g_{ab}z^{a}_{m}z^{b}_{n}=\left(\begin{array}{cccc}
	0 & 1 & 0 & 0  \\
	1 & 0 & 0 & 0 \\
        0 & 0 & 0 & -1 \\
        0 & 0 & -1 & 0
	\end{array}\right).
\end{equation}
The directional derivatives are defined as
\begin{equation}
D=l^{a}\nabla_{a},\quad\quad \Delta=n^{a}\nabla_{a},\quad\quad \delta=m^{a}\nabla_{a},\quad\quad \bar{\delta}=\bar{m}^{a}\nabla_{a},
\end{equation}
whereas the spin coefficients $\{\alpha,\beta,\gamma,\epsilon;\pi,\nu,\mu,\lambda;\kappa,\tau,\sigma,\varrho\}$, which describe the change in the null tetrad from point to point, follow the conventions of \cite{frolov1979newman}. 

From a generic electromagnetic field $F_{ab}$ in a source-free region of $(\mathcal{M},g_{ab})$ one can form the following three complex scalars
\begin{equation}
\Phi_{0}=l^{a}m^{b}F_{ab},\quad\quad\Phi_{1}=\frac{1}{2}(l^{a}n^{b}+\bar{m}^{a}m^{b})F_{ab},\quad\quad \Phi_{2}=\bar{m}^{a}n^{b}F_{ab},
\end{equation}
and, in terms of them, the Maxwell equations in this region become
\begin{eqnarray}
D\Phi_{1}-\bar{\delta}\Phi_{0}&=&(\pi-2\alpha)\Phi_{0}+2\varrho\Phi_{1}-\kappa\Phi_{2},\\
D\Phi_{2}-\bar{\delta}\Phi_{1}&=&-\lambda\Phi_{0}+2\pi\Phi_{1}+(\varrho-2\epsilon)\Phi_{2},\\
\delta\Phi_{1}-\Delta\Phi_{0}&=&(\mu-2\gamma)\Phi_{0}+2\tau\Phi_{1}-\sigma\Phi_{2},\\
\delta\Phi_{2}-\Delta\Phi_{1}&=&-\nu\Phi_{0}+2\mu\Phi_{1}-(\tau-2\beta)\Phi_{2},
\end{eqnarray}
which constitute a coupled system of first-order partial differential equations. In general, it is quite difficult to solve this system before finding a scheme to decouple the equations somehow.

Fortunately, drastic simplifications emerge when we consider a null electromagnetic field adapted to either the null congruence $\Gamma(l)$ or $\Gamma(n)$. In the first case, the solution is called \textit{outgoing} whereas in the latter it is called \textit{ingoing} and, since we are mainly interested in the last case, we write the decomposition as
\begin{equation}\label{nullf}
F^{ab}=\Phi_{0}\bar{m}^{[a}n^{b]}+\bar{\Phi}_{0}m^{[a}n^{b]},
\end{equation}
from which one obtains $\Phi_{1}=\Phi_{2}=0$, and
\begin{eqnarray}\label{null1}
\bar{\delta}\Phi_{0}&=&(2\alpha-\pi)\Phi_{0},\\\label{null2}
\Delta\Phi_{0}&=&(2\gamma-\mu)\Phi_{0},
\end{eqnarray}
together with the condition $\lambda=\nu=0$. The latter is nothing but the statement that the congruence $\Gamma(n)$ must be geodesic and shear-free, as is required by Robinson's theorem \cite{mariot1954champ, robinson1961null}. Also, one easily deduces from equations (\ref{intensity}) and (\ref{nullf}) that the field intensity is given by $E^{2}=2\Phi_{0}\bar{\Phi}_{0}$. In the next section, we present the Kerr congruence in some detail and construct a null electromagnetic field adapted to $\Gamma(n)$ in the Kerr spacetime. Afterward, we develop a simple invariance result that permits us to transport the solution to Minkowski spacetime and to construct the optical metrics in terms of this solution.

\section{The Kerr congruence and its adapted null solution}

\begin{figure}
    \centering
    \includegraphics[width=0.28\linewidth]{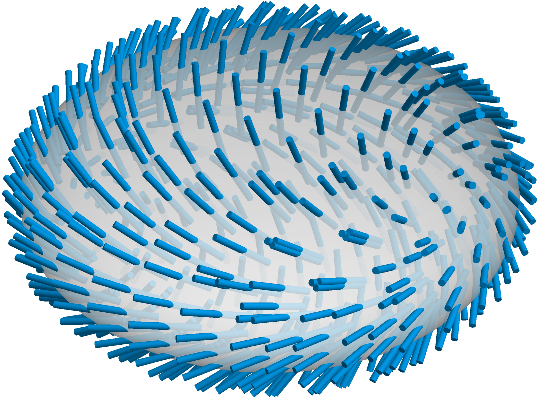}
    \caption{Geometry of the unitary vector field $s^{k}$ in $\mathbb{R}^{3}$. Apart from the disk $x^2+y^2<a^2$ on the equatorial plane, it is well-defined everywhere. Notice that the vector points inwards the confocal ellipsoids and has a reflection symmetry about the equatorial plane.}
    \label{fig:Poynting}
\end{figure}

In the context of general relativity, the spacetime geometry surrounding a rotating black hole with mass $m$ and angular momentum $J=am$ is described by the Kerr metric. In ``cartesian'' coordinates $(t,x,y,z)$, the line element reads as \cite{Stephani_2003,visser2008kerr}:
\begin{eqnarray}\nonumber
ds^{2}&=&dt^{2}-dx^{2}-dy^{2}-dz^{2}\\\label{Cartesian}
&-&\frac{r_{s}r^{3}}{r^{4}+a^{2}z^{2}}\left[dt+\frac{r(xdx+ydy)}{a^{2}+r^{2}}+\frac{a(ydx-xdy)}{a^{2}+r^{2}}+\frac{z}{r}\right]^{2},
\end{eqnarray}
where $r_{s}=2m$ is the Schwarzschild radius and the surfaces of constant $r$ are confocal ellipsoids of revolution about the $z$-axis, being implicitly determined by
\begin{equation}
\frac{x^{2}+y^{2}}{r^{2}+a^{2}}+\frac{z^{2}}{r^{2}}=1.
\end{equation}
We notice that, in the equatorial plane ($z=0$), the function $r(x,y,z)$ becomes complex for $x^{2}+y^{2}<a^{2}$ and vanishes on top of the ring $x^{2}+y^{2}=a^{2}$. In its Kerr-Schild form, the infinitesimal line element (\ref{Cartesian}) gives the following metric tensor:
\begin{equation}
g_{ab}=\mathring{g}_{ab}-\frac{r_s r^{3}}{r^{4}+a^2z^{2}}n_{a}n_{b},
\end{equation}
where $\mathring{g}_{ab}$ is Minkowski metric in Cartesian coordinates and $n^{a}\equiv (1,s^{k})$ is defined in terms of the unit spacelike vector:
\begin{equation}
s^{k}=-\left(\frac{rx+ay}{r^{2}+a^{2}},\frac{ry-ax}{r^{2}+a^{2}},\frac{z}{r}\right),\quad\quad\quad\delta_{ij}s^{i}s^{j}=1.
\end{equation}
A direct calculation then reveals that
\begin{equation}
\mathring{g}_{ab}n^{a}n^{b}=g_{ab}n^{a}n^{b}=0.
\end{equation}
In other words, the congruence spanned by $n^{a}$ is light-like in both the Minkowski and Kerr geometries. Henceforth, the congruence $\Gamma(n)$ will be called the \textit{Kerr congruence}. Its geometric skeleton is sketched in Figure (\ref{fig:Poynting}) and it turns out that this congruence is geodesic and shear-free with respect to both geometries.

In order to construct an ingoing electromagnetic field adapted to $\Gamma(n)$ in Minkowski spacetime $(\mathcal{M}, \mathring{g}_{ab})$, it is highly instructive to first construct it in Kerr spacetime $(\mathcal{M},g_{ab})$ itself. However, since the solution in Cartesian coordinates looks highly involved, it is convenient to work in Boyer-Lindquist coordinates. In these new coordinates $(t,r,\theta,\phi)$, the line element (\ref{Cartesian}) reads as follows
\begin{eqnarray}\nonumber
ds^{2}&=&\left(1-\frac{r_{s}r}{\Sigma}\right)dt^{2}+2\frac{r_{s}ra\mbox{sin}^{2}\theta}{\Sigma}dt d\phi\\\label{BLLE}
&-&\frac{\Sigma}{\Delta}dr^{2}-\Sigma d\theta^{2}-\left(r^{2}+a^{2}+\frac{r_{s}ra^{2}\mbox{sin}^{2}\theta}{\Sigma}\right)\mbox{sin}^{2}\theta d\phi^{2},
\end{eqnarray}
where
\begin{equation}
\Sigma=r^{2}+a^{2}\mbox{cos}^{2}\theta,\quad\quad\quad\Delta=r^{2}-r_{s}r+a^{2}=(r-r_{+})(r-r_{-}),
\end{equation}
with $r_{\pm}=m\pm\sqrt{m^{2}-a^{2}}$. Here, the hypersurface $r=r_+$ corresponds to the outer event horizon, whereas the hypersurface $r=r_{-}$ corresponds to the inner event horizon. Furthermore, the infinite redshift surface is given by $r_{{\rm erg}}=m+\sqrt{m^{2}-a^{2}\cos^2\theta}$, and the spacetime region between the hypersurfaces $r=r_{+}$ and $r=r_{{\rm erg}}$ is called the \textit{ergoregion}. 

It is even more convenient to rewrite the line element (\ref{BLLE}) in its corresponding Kerr-Schild form. In terms of metric tensors, we have
\begin{equation}\label{Kerrst}
g_{ab}=\mathring{g}_{ab}-H_{{\rm Kerr}}n_{a}n_{b}, \quad\quad\quad H_{{\rm Kerr}}=\frac{4r_{s}r\Sigma}{\Delta^{2}},
\end{equation}
where, now
\begin{equation}\label{back}
\mathring{g}_{ab}= \left(\begin{array}{cccc}
1& r_{s}r/\Delta&0 &0\\
r_{s}r/\Delta&-\Sigma(\Delta-r_{s}r)/\Delta^{2}&0 &-r_{s}ra\sin^{2}\theta/\Delta\\
0 &0 &-\Sigma &0 \\
0&-r_{s}ra\sin^{2}\theta/\Delta&0 &-(r^{2}+a^{2})\sin^{2}\theta\\
\end{array}\right),
\end{equation}
is simply Minkowski metric in disguise and the null covector has the following components
\begin{equation}\label{nullcov}
n_{a}=\frac{\Delta}{2\Sigma}\left(1,\frac{\Sigma}{\Delta},0,-a\mbox{sin}^{2}\theta\right).
\end{equation}
By properly inverting equation (\ref{back}) and defining the corresponding vector field $n^{a}=\mathring{g}^{ab}n_{b}$, we construct the so-called Kinnersley tetrad as \cite{kinnersley1969type}
\begin{eqnarray*}
l^{a}&=&\left(\frac{r^{2}+a^{2}}{\Delta},1,0,\frac{a}{\Delta}\right),\\
n^{a}&=&\frac{1}{2\Sigma}\left(r^{2}+a^{2},-\Delta,0,a\right),\\
m^{a}&=&\frac{1}{\sqrt{2}(r+ia\mbox{cos}\theta)}\left(ia\mbox{sin}\theta,0,1,\frac{i}{\mbox{sin}\theta}\right),\\
\bar{m}^{a}&=&\frac{1}{\sqrt{2}(r-ia\mbox{cos}\theta)}\left(-ia\mbox{sin}\theta,0,1,-\frac{i}{\mbox{sin}\theta}\right),
\end{eqnarray*}
which satisfy the normalization conditions given by equation (\ref{norm}) with respect to the Kerr metric by construction. It is worth mentioning that, in this coordinate system, $n^{a}$ is not well defined when $r=0$ and $\theta=\pi/2$ and that this is a manifestation of the disk singularity mentioned before.

The null electromagnetic bivector adapted to the ingoing null congruence $\Gamma(n)$ is assumed to be of the form (\ref{nullf}) as before, and the relevant spin coefficients read as
\begin{eqnarray}
\fl\qquad
\alpha=-\frac{(r+ia\mbox{cos}\theta)\mbox{cos}\theta-2ia}{2\sqrt{2}(r-ia\mbox{cos}\theta)^{2}\mbox{sin}\theta},\quad\quad\gamma=-\frac{\Delta(r+ia\mbox{cos}\theta)-(r-r_{s}/2)\Sigma}{2\Sigma^{2}},\\[1ex]
\fl\qquad
\mu=-\frac{\Delta}{2\Sigma (r-ia\mbox{cos}\theta)},\quad\quad\pi=\frac{ia\mbox{sin}\theta}{\sqrt{2}(r-ia\mbox{cos}\theta)^{2}}.
\end{eqnarray}
Plugging the above quantities in equations (\ref{null1}) and (\ref{null2}) we get a system of first-order linear inhomogeneous differential 
equations for the function $\Phi_{0}$. It turns out that the solution of this system is known and is given by
\begin{equation}\label{Phizero}
\Phi_{0}=\frac{r-ia\mbox{cos}\theta}{\mbox{sin}\theta\Delta}e^{G(\psi_{1},\psi_{2})}, 
\end{equation}
where $G(\psi_{1},\psi_{2})$ is an arbitrary function and
\begin{eqnarray}\label{psi1}
\psi_{1}&=&t+r+m\ln\left|\Delta\right|+\frac{m^{2}}{\sqrt{m^{2}-a^{2}}}\ln\left|\frac{r-r_{+}}{r-r_{-}}\right|-ia\cos\theta,\\\label{psi2}
\psi_{2}&=&\phi+\frac{a}{2\sqrt{m^{2}-a^{2}}}\ln\left|\frac{r-r_{+}}{r-r_{-}}\right|+i\ln\left|\frac{1-\cos\theta}{\sin\theta}\right|.
\end{eqnarray}
This class of null solutions has been discussed extensively in references \cite{pelykh2018class, pelykh2018null} and is a generalization of those solutions first obtained in the seminal paper by Teukolsky on the perturbations about a rotating black hole \cite{teukolsky1973perturbations}.

\section{Transporting the solution to Minkowski spacetime}
In this section, we briefly discuss an invariance result that allows us to transport the solution provided by equations (\ref{Phizero}), (\ref{psi1}) and (\ref{psi2}) to Minkowski spacetime $(\mathcal{M},\mathring{g}_{ab})$. The argument is quite general and proceeds in two simple steps:
\begin{enumerate}
\item{Suppose we have a null bivector $F^{ab}$ in a target spacetime $(\mathcal{M},g_{ab})$ and let its PND be defined by the algebraic relations
\begin{equation}
F^{ab}n_{b}=0,\quad\quad \star F^{ab}n_{b}=0,\quad\quad n^{a}=g^{ab}n_{b},\quad\quad n^{a}n_{a}=0.
\end{equation}
Clearly, the algebraic invariants (\ref{invs}) identically vanish as a consequence of our assumption. Now, assume that a given background $\mathring{g}_{ab}$ metric and the target metric $g_{ab}$ are related by a disformal transformation of the type
\begin{equation}
\mathring{g}_{ab}=g_{ab}+fn_{a}n_{b},\quad\quad\quad \mathring{g}^{ab}=g^{ab}-fn^{a}n^{b},
\end{equation}
for some scalar $f$. A direct calculation reveals that the transformation is volume-preserving i.e., the determinants coincide. Furthermore, a careful inspection of the Hodge star operator $\mathring{\star}$ with respect to $\mathring{g}_{ab}$ gives
\begin{equation}
\mathring{\star}{F}^{ab}=\star{F}^{ab},\quad\quad\quad \mathring{\star}{F}_{ab}=\star{F}_{ab}.
\end{equation}
In other words: if the bivector $F^{ab}$ is null with respect to the target metric $g_{ab}$, it is also null with respect to the background metric $\mathring{g}_{ab}$ and the result is independent on the function $f$. The conclusion is straightforward since the PND coincides and may be alternatively obtained by calculating the two independent invariants provided by equations (\ref{invs}).
}
\item{Suppose we have a null bivector $F^{ab}$ in a target spacetime $(\mathcal{M},g_{ab})$ and let its PND and the background metric be defined as before. Furthermore, assume that the bivector is an exact solution of Maxwell equations in vacuum i.e.
\begin{equation}
\frac{1}{\sqrt{-g}}(\sqrt{-g}F^{ab})_{,b}=0,\quad\quad\quad \frac{1}{\sqrt{-g}}(\sqrt{-g}\star F^{ab})_{,b}=0.
\end{equation}
Now, since the transformation of the metric is volume-preserving and the Hodge duals coincide, it is clear that the same bivector will be automatically a solution in the background space $(\mathcal{M},\mathring{g}_{ab})$ as well.
}
\end{enumerate}
The results discussed here constitute a particular instance of the so-called disformal invariance of Maxwell's field equations in vacuum. Indeed, this type of invariance is not restricted to null fields, but it is also valid for regular ones. The full demonstration together with some algebraic and group properties of this class of metric transformations can be seen in  \cite{goulart2013disformal,goulart2021space}.

\section{The rotating black hole analogue}

In this section, we collect the results discussed so far to obtain a family of optical metrics for \textit{photons} which qualitatively mimic the behavior of rotating black holes in the context of general relativity. We shall see that the family is characterized by three parameters and exhibits special hypersurfaces which would correspond to the outer ergosurface and outer event horizon as they appear in the Kerr metric. For the sake of concreteness, we start with Minkowski spacetime $(\mathcal{M},\mathring{g}_{ab})$ in coordinates $x^{a}=(t,r,\theta,\phi)$ such that the metric tensor and the Kerr congruence have components provided by equations (\ref{back}) and (\ref{nullcov}). We then restrict ourselves to the simplest possible null electromagnetic field adapted to $\Gamma(n)$ by putting $G(\psi_{1},\psi_{2})=0$ in equation (\ref{Phizero}). With these assumptions, we get
\begin{equation}\label{optst}
\Phi_{0}=\frac{r-ia\cos\theta}{\sin\theta\Delta}\quad\quad\rightarrow\quad\quad H_{{\rm opt}}=\frac{2\chi\Sigma}{\sin^{2}\theta\Delta^{2}},
\end{equation}
and plugging this into equations (\ref{effmets}) together with equation (\ref{nullcov}), we obtain the covariant optical metric

\begin{equation}\label{eq:opt_met_case1}
\begin{array}{l}
\tilde g_{00}=1-\frac{\chi}{2\sin^2\theta\Sigma},\qquad
\tilde g_{01}=\frac{1}{\Delta}\left(r_{s}r-\frac{\chi}{2\sin^2\theta}\right),\qquad\tilde g_{03}=\frac{\chi a}{2\Sigma},\\[1ex]
\tilde g_{11}=-\frac{1}{\Delta^{2}}\left[\Sigma(\Delta-r_{s}r) + \frac{\chi \Sigma}{2 \sin^2\theta}\right],\qquad \tilde g_{13}=-\frac{r_{s}ra\sin^2\theta}{\Delta} + \frac{\chi a }{2 \Delta},\\[1ex]
\tilde g_{22}=-\Sigma,\qquad \tilde g_{33}=-\left(r^{2}+a^{2}+\frac{\chi a^2}{2\Sigma}\right)\sin^{2}\theta.
\end{array}
\end{equation}
Hence, one has the presence of the three parameters $\chi$, $m$ and $a$. Furthermore, due to its simple Kerr-Schild form, it is easily seen that the determinant behaves as
\begin{equation}\label{det}
\mbox{det}(\tilde{g}_{ab}) = -\Sigma^2\sin^2\theta. 
\end{equation}
In other words, the latter is the same as the determinant of the background metric $\mathring{g}_{ab}$.

\subsection{Singularity structure}

From equation (\ref{det}) one realizes that the optical metric is degenerate at $\Sigma=0$ or at the semi-axis $\theta=0$ and $\theta=\pi$. To reveal that these coordinate singularities are actually true curvature singularities, we compute the components of the contravariant optical metric, to get 
\begin{equation}\label{eq:inv_opt_met_case1}
\begin{array}{l}
\tilde g^{00}= -\frac{1}{2}\frac{(a^2+r^2)^2\chi + 2\sin^2\theta[(\Delta-2Mr)(a^2+r^2)\Sigma - 4M^2a^2r^2\sin^2\theta]}{\sin^2\theta\Sigma\Delta^2},\\[1ex]
\tilde g^{01}=-\frac{1}{2}\frac{(\chi-4Mr\sin^2\theta)(a^2+r^2)}{\sin^2\theta\,\Delta\Sigma},\qquad\tilde g^{03}=\frac{1}{2}\frac{[\chi(r^2+a^2) - 8M^2r^2\sin^2\theta]a}{\sin^2\theta \,\Delta^2 \Sigma},\\[1ex]
\tilde g^{11}=-\frac{1}{\Sigma}\left(r^2 + a^2 -\frac{\chi}{2\sin^2\theta}\right),\qquad \tilde g^{13}=-\frac{1}{2}\frac{(\chi-4Mr\sin^2\theta)a}{\sin^2\theta\, \Delta\Sigma},\\[1ex]
\tilde g^{22}=-\frac{1}{\Sigma},\qquad \tilde g^{33}=-\frac{1}{2}\frac{4Mr(2Mr-\Sigma) + 2\Sigma\Delta - \chi a^2}{\sin^2\theta\,\Delta^2\Sigma}.
\end{array}
\end{equation}
Some computational symbolic manipulation then shows that the Ricci scalar identically vanishes, but the Kretschmann invariant is given by
\begin{equation}
\label{optical_krets}
R^{abcd}R_{abcd}=\frac{2\chi^2(7a^4\cos^4\theta - 34a^2 r^2 \cos^2\theta + 7r^4)}{\Sigma^6\sin^4\theta}.
\end{equation}
Clearly, the latter also vanishes when one deletes the non-linearity parameter, as expected. However, if $\chi$ is finite, the invariant will blow up precisely when $\Sigma=0$ or at the semi-axis $\theta=0$ and $\theta=\pi$, showing that these regions are true singularities of the optical metrics. When written in Cartesian coordinates, these singularities precisely coincide with the axis of rotation and the ring with radius $a$ in the equatorial plane.

\subsection{Special surfaces}

\begin{figure}
    \centering
    \includegraphics[width=0.35\linewidth]{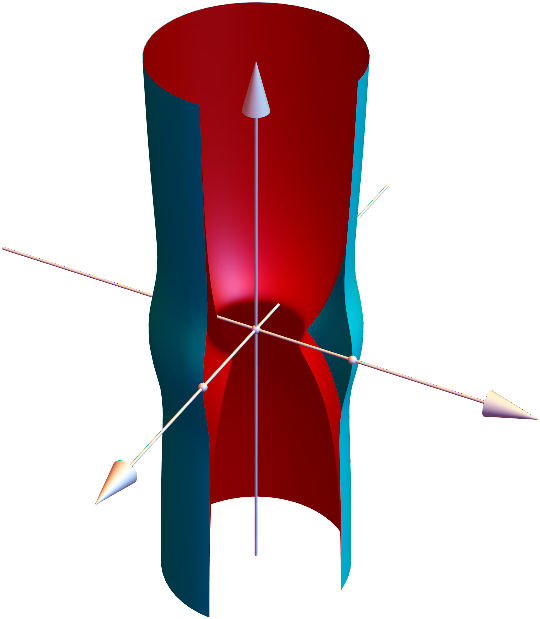}
    \includegraphics[width=0.35\linewidth]{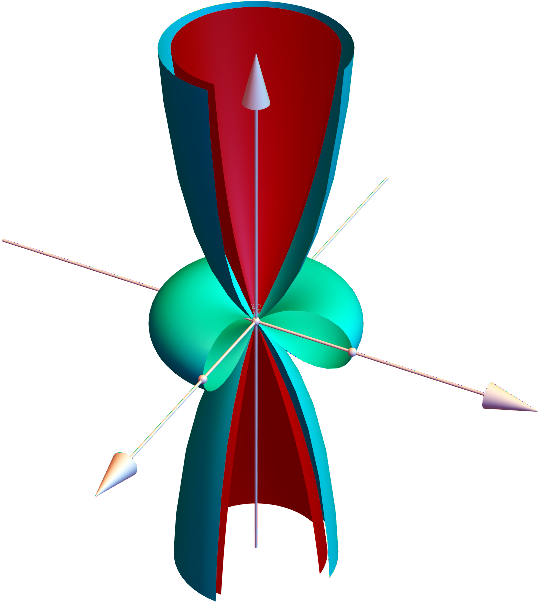}
    \caption{Internal anatomy of the maximum (in red) and ergo (in blue) hypersurfaces: (i) slowly rotating case with $2\chi>a^2$ (left) and (ii) rapidly rotating case with $2\chi\leq a^2$ (right). We have added a cut to reveal the internal structures in more detail. In both cases, the intersection of the ergosurface with the equatorial plane occurs for $r=\sqrt{\chi/2}$.}
    \label{fig:ergo_ini}
\end{figure}

The determination of the optical ergosurface is done by finding the space-time loci where photons are infinitely redshifted with respect to observers at spatial infinity. Another way of saying this is the following: suppose we have an observer that tries to “stand still” at a fixed (spatial) point in the coordinate system, as described in \cite{visser2008kerr}. This is the same as to require the worldline condition
\begin{equation}
x^{a}(t)=(t,r_{0},\theta_{0},\phi_{0})
\end{equation}
for some numbers $r_{0}$, $\theta_{0}$ and $\theta_{0}$. Clearly, this worldline must be timelike and this is possible only when $\tilde{g}_{00}>0$. However, setting the component $\tilde g_{00}=0$, we get
\begin{equation}
\label{eq_redsh}
r_{{\rm erg}}=\frac{1}{2}\mbox{csc}(\theta)
\sqrt{2\chi-a^2\mbox{sin}^{2}(2\theta)},
\end{equation}
which is the analog of the ergosurface in Kerr spacetime. In other words, for $r_{0}<r_{{\rm erg}}$, it is impossible for the observer to stand still and the light cone provided by the optical metric has a direction that is tangent to his worldline. The axial symmetry of this ergosurface is evident, but the corresponding surface will not be compact as in the Kerr metric. Indeed, if the parameter $2\chi>a^2$, then for all $\theta$ there is one $r>0$ satisfying equation (\ref{eq_redsh}). In this case, the intersection of the ergosurface with $3$-space will be homeomorphic to a cylinder. Conversely, when $2\chi\leq a^2$, then there is an interval of $\theta$ for which equation (\ref{eq_redsh}) does not have real roots for $r$. In this case, the corresponding surface will have a more complex shape containing self-intersections. In other words, when $\chi$ is fixed, the ergoregion type depends on whether the analog black hole has slow or fast rotation but, perhaps surprisingly, all ergosurfaces pass through the ring with radius $\sqrt{\chi/2}$ in the equatorial plane $\theta=\pi/2$, which does not depend on the angular momentum. The shape of the possible ergoregions is depicted by the external surfaces (in blue) in figure \ref{fig:ergo_ini}, for both cases of a slow and fast rotating analog black hole.

In order to show that an effective event horizon exists for every member of the family of optical metrics, we proceed as follows. Consider a $3$-dimensional surface defined by an equation $r_{{\rm hor}}=r(\theta)$ and let the other three coordinates $(t,\theta,\phi)$ run over their respective ranges. Clearly, this hypersurface inherits the axial symmetry of the solution and we can compute the optical metric restricted to it, that is
\begin{equation}\label{induced}
\tilde{h}_{\alpha\beta}= \left(\begin{array}{ccc}
\tilde{g}_{00}& \tilde{g}_{01}r'& \tilde{g}_{03} \\
\tilde{g}_{01}r'&\tilde{g}_{11}r'^{2}+\tilde{g}_{22}&\tilde{g}_{13}r' \\
\tilde{g}_{03} &\tilde{g}_{13}r' &\tilde{g}_{33}  \\
\end{array}\right),
\end{equation}
where $r'=dr/d\theta$, for conciseness. The idea is to show that there is one null hypersurface that separates the spacetime into two disjoint regions. To do so, a simple calculation shows that the induced optical metric has the following determinant
\begin{equation}\label{pulldet}
\mbox{det}(\tilde{h}_{\alpha\beta})=\Sigma\sin^2\theta\left(r'^{2}+r^{2}+a^{2}-\frac{\chi}{2}\csc^2\theta\right).
\end{equation}
In general relativity, a null hypersurface is a hypersurface whose normal co-vector at every point is light-like. Another way of saying this is that the pullback (restriction) of the ambient metric onto the hypersurface is degenerate. Therefore, from equation (\ref{pulldet}), one obtains the nonlinear ordinary differential equation
\begin{equation}\label{nledo}
r'^{2}=\frac{\chi}{2}\mbox{csc}^{2}\theta-r^{2}-a^{2},
\end{equation}
which only makes sense if $r\leq r_{{\rm max}}$, where
\begin{equation}\label{maxhyp}
r_{{\rm max}}=\sqrt{\frac{\chi}{2}\csc^{2}\theta-a^{2}}
\end{equation}
defines what we call henceforth the \textit{maximum hypersurface}. It is easily seen from equations (\ref{eq_redsh}) and (\ref{maxhyp}) that the maximum hypersurface is inside the ergosurface ($r_{{\rm max}}<r_{{\rm erg}}$) and that they almost coincide for $\theta\rightarrow 0$ and $\theta\rightarrow \pi$. In figure (\ref{fig:ergo_ini}) we have plotted these surfaces (in red) for different values of the parameters to elucidate the problem.

\begin{figure}
    \centering
    \includegraphics[width=0.35\linewidth]{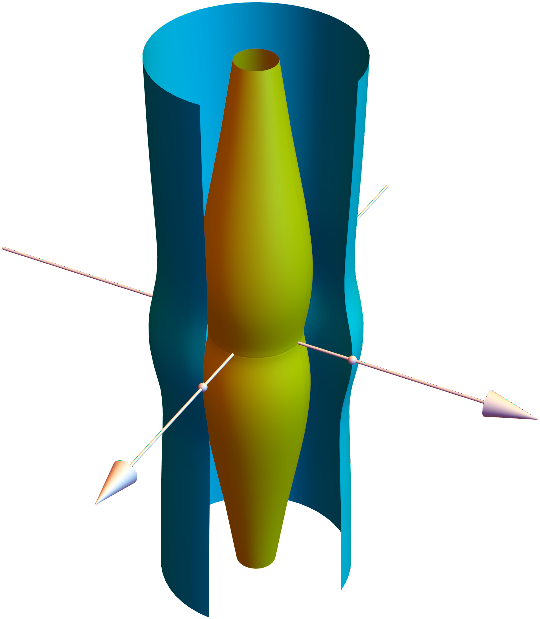}
    \includegraphics[width=0.35\linewidth]{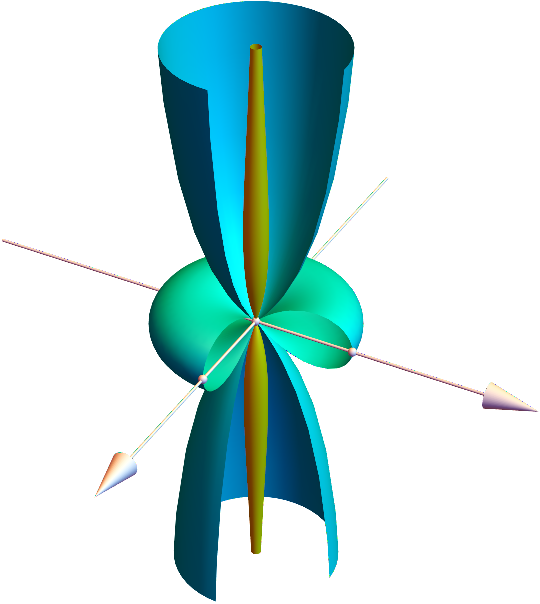}
    \caption{Ergosurface (in blue) and event horizon (in yellow): (i) slowly rotating case with $\chi>2a^{2}$ (left) and (ii) rapidly rotating case with $\chi\leq2a^{2}$ (right). Notice that for sufficiently large rotation, the horizon intersects itself at the ``origin''.}
    \label{fig:ergo_hor}
\end{figure}

Thus, we must solve the first-order nonlinear ordinary differential equation given by equation (\ref{nledo}) subjected to the condition $r\leq r_{max}$. Unfortunately, this equation cannot be solved analytically. However, using a computational resource in the framework of the characteristics method \cite{John1982} to guarantee the existence and uniqueness of solutions, we could find all the candidates for the analog event horizon. Each valid initial condition of equation (\ref{nledo}) gives a curve in the interior of the maximum hypersurface and different initial conditions starting over the surface do not intersect with each other. Therefore, the collection of curves, representing null hypersurfaces, forms a foliation of the inner region of the maximum hypersurface. Although all of them are null, there is only one that separates the spacetime into two disjoint regions and, besides, is the most external possible one. All the other null hypersurfaces whose initial condition is given on top of the maximum hypersurface do not satisfy the splitting condition. It should be clear that photons emitted inside these regions cannot traverse them, thus defining a boundary beyond which events cannot affect an observer located at large distances. The shape of the analog event horizon depends on the choice of $\chi$ and $a$, and it is depicted in figure \ref{fig:ergo_hor} by the innermost (yellow) surfaces. In particular, the transition in topology occurs when $\chi=2a^2$, when the ``throat'' of the event horizon (the value of $r$ at the equator plane) shrinks achieving the ring singularity and intersecting itself. Thus, for the slowly rotating case ($\chi>2a^2$) the size of the throat is non-zero, while for the rapidly rotating one ($\chi\leq2a^2$) the throat size vanishes.

\subsection{Special slice}

It is clear from the above discussion that the essential difference between the original Kerr metric (\ref{Kerrst}) and the optical metric engendered by equation (\ref{optst}) resides in the scalars $H_{{\rm Kerr}}$ and $H_{{\rm opt}}$. Indeed,  although the background metric $\mathring{g}_{ab}$ and the geodesic shear-free congruence spanned by $n^{a}$ are precisely the same in both cases, $H_{{\rm Kerr}}$ is constrained to satisfy Einstein field equations, whereas $H_{{\rm opt}}$ is constrained to satisfy Maxwell field equations. A measure of this difference is provided by the ratio:
\begin{equation}
\frac{H_{\rm opt}}{H_{\rm Kerr}}=\frac{\chi}{2 r_{s}\mbox{sin}^{2}\theta r}.
\end{equation}
From this, one realizes that the geometries drastically differ as we approach the axis of rotation. Also, one sees that for some given $\theta$ not on the axis of rotation, the optical metric decays faster than the Kerr metric. In other words, the optical metric approaches the Minkowski metric faster, thus becoming asymptotically flat first. This phenomenon is expected since the optical metric involves the square of the background field intensity $E(x)$ which decays as $1/r$ in the radiation zone. 

A natural question arises here. Is there a particular submanifold where the two metrics coincide? The answer is yes and is obtained by assuming:
\begin{equation}
\frac{H_{\rm opt}}{H_{\rm Kerr}}=1\quad\quad\rightarrow\quad\quad r_{\rm slice}=\frac{\chi}{2 r_{s}\mbox{sin}^{2}\theta}.
\end{equation}
Clearly, on top of this embedded hypersurface, the Kerr metric and the optical metric describe exactly the same line element. This result is in contrast with the result that the Kerr equator can (in principle) be exactly simulated by an acoustic analog based on a vortex flow with a very specific equation of state and
subjected to a very specific external force \cite{visser2005vortex}. Interestingly, in our case, the equation describing this special slice does not depend on the angular momentum of the analogue black hole. 

\section{Concluding remarks}

We briefly review the main steps for the derivation of an effective optical metric for light rays in the realm of nonlinear electrodynamics. Particularly, when both electromagnetic invariants vanish, we show that there is a principal null direction of the electromagnetic tensor that reduces the optical metric to a Kerr-Schild-like form. Such simplification is even more transparent when using the NP formalism.

By using the Kerr congruence and its adapted null solution \cite{teukolsky1973perturbations,pelykh2018class,pelykh2018null}, we demonstrate that such null electromagnetic configurations can be directly transposed to the context of Minkowski spacetime. Then, we investigate the properties of the corresponding optical metric, showing that it represents an interesting analogue for rotating black holes. Firstly, we found that this metric presents a ring singularity similar to the Kerr black hole. Secondly, it defines a regular ergosurface and a regular event horizon, whose specific shapes depend on the angular momentum of the black hole. Curiously enough, both hypersurfaces do not depend on the mass of the analog black hole, which is different from the Kerr metric case. Finally, we show that there exists a three-dimensional slice of the effective spacetime that has exactly the same geometry of a slice of the Kerr metric.

Contrary to some previous cases investigated in the literature \cite{Novello_2003,DeLorenci2003, Bittencourt2014,PhysRevA.104.043523}, this optical metric is safe from coordinate system issues (see the reviews \cite{barcelo2005,novello2002artificial}). Indeed, it can be checked that the characteristic polynomial is always hyperbolic, the maximal speed of propagation does not exceed the ordinary velocity of light and it is regular almost everywhere, including both the ergosurface and the event horizon. This result is in contrast with several other models that may suffer from troubles when trying to cross the event horizon, violation of the geometric limit, divergences of the background fields, lack of hyperbolicity, signature transition issues, among others.

Finally, for further investigation, it would be interesting to investigate the viability of building such a configuration in the laboratory by using the equivalence between nonlinear electrodynamics and nonlinear material media, where a given nonlinear Lagrangian can be mapped into constitutive relations between the field strengths and their corresponding excitations.

\section*{Acknowledgements}
EB thanks CNPq for the financial support (grant N.\ 305217/2022-4). We are in debt to L.S. Ruiz for valuable comments on a previous version of the paper.

\section*{References}
\bibliography{refs.bib}

\providecommand{\newblock}{}
\begin{thebibliography}{10}
\expandafter\ifx\csname url\endcsname\relax
  \def\url#1{{\tt #1}}\fi
\expandafter\ifx\csname urlprefix\endcsname\relax\def\urlprefix{URL }\fi
\providecommand{\eprint}[2][]{\url{#2}}

\bibitem{Kerr_1963}
Kerr R~P 1963 {\em Phys. Rev. Lett.\/} {\bf 11}(5) 237--238 \urlprefix\url{https://link.aps.org/doi/10.1103/PhysRevLett.11.237}

\bibitem{kerr1965gravitational}
Kerr R~P 1965 {\em Quasi-Stellar Sources and Gravitational Collapse\/}  99

\bibitem{barcelo2005}
Barcel\'o C, Liberati S and Visser M 2005 {\em Living Reviews in Relativity\/} {\bf 8} 12

\bibitem{novello2002artificial}
Novello M, Visser M and Volovik G 2002 {\em Artificial Black Holes\/} (World Scientific) ISBN 9789810248079 \urlprefix\url{https://books.google.com.br/books?id=-tyXuduShHUC}

\bibitem{visser2005vortex}
Visser M and Weinfurtner S 2005 {\em Classical and Quantum Gravity\/} {\bf 22} 2493

\bibitem{giacomelli2017rotating}
Giacomelli L and Liberati S 2017 {\em Physical Review D\/} {\bf 96} 064014

\bibitem{jacobson1998event}
Jacobson T and Volovik G 1998 {\em Physical Review D\/} {\bf 58} 064021

\bibitem{baak2023petrov}
Baak S~S, Datta S and Fischer U~R 2023 {\em Classical and Quantum Gravity\/} {\bf 40} 215001

\bibitem{Novello2000}
Novello M, De~Lorenci V~A, Salim J~M and Klippert R 2000 {\em Phys. Rev. D\/} {\bf 61}(4) 045001 \urlprefix\url{https://link.aps.org/doi/10.1103/PhysRevD.61.045001}

\bibitem{DELORENCI2000134}
{De Lorenci} V, Klippert R, Novello M and Salim J 2000 {\em Physics Letters B\/} {\bf 482} 134--140 ISSN 0370-2693 \urlprefix\url{https://www.sciencedirect.com/science/article/pii/S0370269300005220}

\bibitem{DUNNE_2005}
Dunne G~V 2005 {\em Heisenberg–Euler Effective Lagragians: Basics and Extensions\/} (World Scientific) p 445–522 \urlprefix\url{http://dx.doi.org/10.1142/9789812775344_0014}

\bibitem{breton2007nonlinear}
Breton N and Garcia-Salcedo R 2007 Nonlinear electrodynamics and black holes (\textit{Preprint} \eprint{hep-th/0702008})

\bibitem{Sorokin_2022}
Sorokin D~P 2022 {\em Fortschritte der Physik\/} {\bf 70} ISSN 1521-3978 \urlprefix\url{http://dx.doi.org/10.1002/prop.202200092}

\bibitem{alam2022review}
Alam Y~F and Behne A 2022 Review of born-infeld electrodynamics (\textit{Preprint} \eprint{2111.08657})

\bibitem{Chandra1983}
Chandrasekhar S 1983 {\em The Mathematical Theory of Black Holes\/} 1st ed (Oxford University Press)

\bibitem{Stephani_2003}
Stephani H, Kramer D, MacCallum M, Hoenselaers C and Herlt E 2003 {\em Exact Solutions of Einstein’s Field Equations\/} 2nd ed Cambridge Monographs on Mathematical Physics (Cambridge University Press)

\bibitem{visser2008kerr}
Visser M 2008 The kerr spacetime: A brief introduction (\textit{Preprint} \eprint{0706.0622})

\bibitem{teukolsky1973perturbations}
Teukolsky S~A 1973 {\em Astrophysical Journal, Vol. 185, pp. 635-648 (1973)\/} {\bf 185} 635--648

\bibitem{pelykh2018class}
Pelykh V and Taistra Y 2018 {\em Journal of Mathematical Sciences\/} {\bf 229}

\bibitem{pelykh2018null}
Pelykh V and Taistra Y 2018 {\em Math. Model. Comput\/} {\bf 5}

\bibitem{GoulartdeOliveiraCosta:2009pr}
Goulart~de Oliveira~Costa E and Esteban Perez~Bergliaffa S 2009 {\em Class. Quant. Grav.\/} {\bf 26} 135015 (\textit{Preprint} \eprint{0905.3673})

\bibitem{Abalos:2015gha}
Abalos F, Carrasco F, Goulart E and Reula O 2015 {\em Phys. Rev. D\/} {\bf 92} 084024 (\textit{Preprint} \eprint{1507.02262})

\bibitem{Goulart:2021uzr}
Goulart E and Bergliaffa S~E~P 2022 {\em Phys. Rev. D\/} {\bf 105} 024021 (\textit{Preprint} \eprint{2112.07078})

\bibitem{Goulart_2024}
Goulart E and Bittencourt E 2024 {\em Classical and Quantum Gravity\/} {\bf 41} 195026

\bibitem{frolov1979newman}
Frolov V 1979 The newman-penrose method in the theory of general relativity {\em Problems in the general theory of relativity and theory of group representations\/} (Springer) pp 73--185

\bibitem{mariot1954champ}
Mariot L 1954 {\em Comptes Rendus Academie des Sciences (serie non specifiee)\/} {\bf 238} 2055--2056

\bibitem{robinson1961null}
Robinson I 1961 {\em Journal of Mathematical Physics\/} {\bf 2} 290--291

\bibitem{kinnersley1969type}
Kinnersley W 1969 {\em Journal of Mathematical Physics\/} {\bf 10} 1195--1203

\bibitem{goulart2013disformal}
Goulart E and Falciano F 2013 {\em Classical and Quantum Gravity\/} {\bf 30} 155020

\bibitem{goulart2021space}
Goulart {\'E} and Bittencourt E 2021 {\em Classical and Quantum Gravity\/} {\bf 38} 145029

\bibitem{John1982}
John F 1982 {\em Partial Differential Equations\/} 4th ed (Cambridge University Press)

\bibitem{Novello_2003}
Novello M, Bergliaffa S~P, Salim J, Lorenci V~A~D and Klippert R 2003 {\em Classical and Quantum Gravity\/} {\bf 20} 859 \urlprefix\url{https://dx.doi.org/10.1088/0264-9381/20/5/306}

\bibitem{DeLorenci2003}
De~Lorenci V~A, Klippert R and Obukhov Y~N 2003 {\em Phys. Rev. D\/} {\bf 68}(6) 061502 \urlprefix\url{https://link.aps.org/doi/10.1103/PhysRevD.68.061502}

\bibitem{Bittencourt2014}
Bittencourt E, Lorenci V~A~D, Klippert R, Novello M and Salim J~M 2014 {\em Classical and Quantum Gravity\/} {\bf 31} 145007 \urlprefix\url{https://dx.doi.org/10.1088/0264-9381/31/14/145007}

\bibitem{PhysRevA.104.043523}
Bittencourt E, Klippert R, da~Silva D~R and Goulart E 2021 {\em Phys. Rev. A\/} {\bf 104}(4) 043523

\end{thebibliography}

\end{document}